\documentclass[aps, prl, twocolumn, showpacs, superscriptaddress,longbibliography]{revtex4}
\usepackage{amssymb}
\usepackage{amsmath}
\usepackage{amsfonts}
\usepackage{amsthm}
\usepackage{bbm}
\usepackage{graphicx}

\usepackage{siunitx}
\usepackage{changes}

\def\be{\begin{equation}}
\def\ee{\end{equation}}

\begin{document}

\title{What determines the ultimate precision of a quantum computer?}

\author{Xavier Waintal}
\thanks{xavier.waintal@cea.fr}
\affiliation{Univ.\ Grenoble Alpes, CEA, INAC-Pheliqs, F-38000 Grenoble, France}
\date{\today}

\begin{abstract}
A quantum error correction (QEC) code uses $N_{\rm c}$ quantum bits to construct one "logical" quantum bits of better quality than the original "physical" ones. QEC theory predicts that the failure probability  $p_L$ of logical qubits decreases exponentially with  
$N_{\rm c}$ provided the failure probability $p$ of the physical qubit is below a certain threshold $p<p_{\rm th}$.
In particular QEC theorems imply that the logical qubits can be made arbitrarily precise by simply increasing $N_{\rm c}$. 
In this letter, we search for physical mechanisms that lie outside of the hypothesis of QEC theorems and set a limit $\eta_{\rm L}$ to the precision of the logical qubits (irrespectively of $N_{\rm c}$). 
$\eta_{\rm L}$ directly controls the maximum number of
operations $\propto 1/\eta_{\rm L}^2$ that can be performed before the logical quantum state gets randomized,
hence the depth of the quantum circuits that can be considered. We identify a type of error - silent stabilizer failure - 
as a mechanism responsible for finite $\eta_{\rm L}$ and discuss its possible causes. Using the example of the topological surface code,
we show that a single local event can provoke the failure of the logical qubit, irrespectively of $N_c$.
\end{abstract}

\maketitle

A quantum computer is an analog machine\cite{landauer1994,dyakonov2014}. 
By this statement,  we mean that its internal state is set by complex numbers that vary {\it continuously }.
A direct consequence of being analogue is that a quantum operation can only be performed with a finite precision $\eta$.
As one performs more operations, the overall precision of the machine deteriorates and eventually the quantum state gets randomized. 
Hence finite precision implies that a finite maximum number $N_\#\sim 1/\eta^2$ of meaningful operations can be performed. 
This number - tagged the maximum "depth" of a quantum circuit - is of prime importance to quantum computing. Quantum algorithms generically 
require that the number of operations increases with the size of the problem (number of qubits) so that 
$\eta$ also determines the maximum size of a useful machine. In short, the precision $\eta$ is the prime indicator of the computing power of a quantum computer.
Typical quantum circuits implemented so far report depths values of a few tens for two qubits circuits \cite{omalley2016} down to a few units for circuits with a few qubits \cite{peruzo2014,kandala2017} or more than ten qubits \cite{otterbach2017}. 
An increasing effort is devoted to studying the advantages that could be harnessed from small depth quantum 
computers \cite{moll2017,neill2017,preskill2018}. 
However, the long term goal of quantum computing still remains achieving an exponential acceleration with respect to a classical computer, 
which implies large depths. For instance, the study of \cite{reiher2017} - devoted to the application of quantum algorithms to assist
quantum chemistry calculations - estimates that $N_\# \sim 10^{15}$ operations are needed to achieve chemical precision in a $50$ levels 
calculation ($50$ electronic levels being roughly the maximum size that can be calculated with brute force classical computing). 

Quantum Error Correction (QEC) proposes a solution to the precision problem by building "logical qubits" out of several
($N_c$) physical qubits. The logical qubits are more robust and more precise than the original qubits from which they are built. 
A crucial result of QEC theory is that the 
precision $\eta_{\rm L}$ of the logical qubit improves {\it exponentially} with $N_c$. In other words, QEC allows one to
build logical qubits of {\it unlimited} precision (hence unlimited depth $N_\#$) at an affordable cost in terms of hardware. This is a very strong statement. 
A natural question, that we will try to address in this letter is therefore what ultimately limits the precision $\eta_{\rm L}$ that can be reached by the logical qubits. Can the precision $\eta_{\rm L}$ improve without limits? Within QEC theorems, the answer is yes. 
However these theorems rely on certain hypothesis that some physical mechanisms, such as correlated or leakage errors, do not fulfill.
In this letter, we revisit the QEC construction and seek mechanisms that would set a limit to $\eta_{\rm L}$ irrespectively of the value of $N_c$. We find a class of errors - silent stabilizer failure - that provides a generic mechanism for limiting $\eta_{\rm L}$ to a finite value. 

\begin{figure} 
\includegraphics[width=7cm]{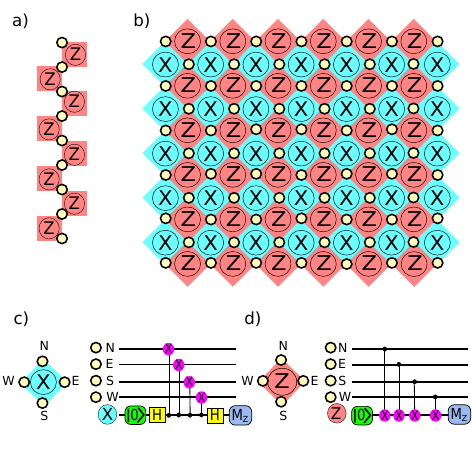} 
\vskip -1cm
\caption{Schematic of a $N_c=9$ repetition code (a) and a $N_c=72$ surface code (b): the small yellow circles indicate the data qubits. The
larger circles with an X or a Z embedded indicate ancilla qubits that are used to measure the stabilizers. The X (Z) ancillas
are used to measure the stabilizer $X_N X_E X_S X_W$ ($Z_N Z_E Z_S Z_W$) where $N,E,S,W$ indicate the data qubit situated North, East, 
South and West with respect to the position of the ancilla. Panel c) and d) indicate the circuit used to measure the X and Z type of stabilizers.
The different pictograms indicate $|0\rangle$: initialization in state $|0\rangle$. H: Hadamard gate $H = (X+Z)/\sqrt{2}$. $M_Z$: measurement of the qubit along the $z$ axis. Small black circle linked to an X: Control-Not 2 qubits gate $C(1,2)= (1+Z_1 +X_2 -Z_1X_2)/2$ which flips the second qubit (X) if the first one is in state $|1\rangle$.
\label{fig:surface} } 
\end{figure}

{\it QEC stabilizer theory in a nutshell.} While some of our statements are general, we will illustrate our arguments with two QEC codes: the surface code \cite{fowler2012b} (the most promising architecture at the moment) and the repetition code \cite{fowler2013} (for pedagogical reasons).  
These two codes belong to the general category of "stabilizer" codes \cite{shor1995,steane1996,laflamme1996,shor1996,divicenzo1996,plenio1997,aharonov1997,knill1998,kitaev1997,fowler2012}. 
QEC uses robust states $|0_L\rangle$ (logical zero) and $|1_L\rangle$ (logical one) to spread out the information of a logical qubit onto $N_c$ physical qubits. A logical state $|\Psi_L\rangle$ of the logical qubit is encoded as,
\begin{equation}
|\Psi_L\rangle = \alpha |0_L\rangle + \beta|1_L\rangle.
\end{equation} 
The states $|0_L\rangle$ and $|1_L\rangle$ are given implicitely by a set of $N_c-1$ observables $g_\alpha$ known as the stabilizers. 
The $g_\alpha$ are product of Pauli matrices $X_i$, $Y_i$ and $Z_i$ that act on qubit $i$. They commute with each others, have $\pm 1$ eigenvalues, and verify: $g_\alpha|\Psi_L\rangle = +1 |\Psi_L\rangle$. 
Figure \ref{fig:surface} shows the stabilizers of the repetition and surface code.
QEC consists in repeatidly measuring all the different stabilizers in order to maintain the state inside the logical subspace.
The set of observed values is called the syndrome. The presence of one or more $-1$ in the syndrome indicates the presence of errors
which must be tracked, see for instance \cite{nielsen2000} for details.
In the repetition code (see Fig.\ref{fig:surface}a),  the logical states $|0_L\rangle$ and $|1_L\rangle$ are simply given by
\begin{equation}
|0_L\rangle = |00\dots 00\rangle \ \ , \ \ |1_L\rangle = |11\dots 11\rangle.
\end{equation}
and the stabilizers $g_i = Z_i Z_{i+1}$ check that one qubit and the next are in the same state. 
If an error $|\Psi_L\rangle\rightarrow X_j|\Psi_L\rangle$ happens, then the syndromes of $g_{j-1}$ and $g_j$ change from $+1$ to $-1$. 
The error is detected and can be corrected by applying $X_j$. 
The repetition code obviously cannot correct $Z_i$ type of errors but its cousin where the "Z" stabilizers are replaced by "X" stabilizers ($g_i = X_i X_{i+1}$) can. Figuratively, the surface code can be seen as the product of these two repetition codes in order to correct for all kind of errors. In the discrete error model, each qubit has a probability $p$ to get an error per clock cycle, and the errors are uncorrelated. The logical qubit remains {\it infinitely precise} except when more than half of the data qubits are corrupted. The probability $p_L$ for logical failure $p_L \sim (p/p_{\rm th})^{N_c/2}$ (repetition code) and $p_L \sim (p/p_{\rm th})^{\sqrt{N_c}/2}$ (surface code, $p_{\rm th}\approx 0.01$) decreases exponentially with $N_c$ provided $p$ is smaller than the threshold $p_{\rm th}$. 

{\it Precision errors for the repetition code.} Let us now study how QEC handles precision (also known as coherent) errors. 
Precision errors include small over/under rotation in the application of a gate, the presence of small unknown terms in the qubit Hamiltonian matrix, the effect of small variation of the energy splitting of a qubit, errors in the direction of a measurement and many other sources. Precision errors reflect the fact that handling a qubit can only be done with a finite precision.
We model this imprecision by applying a  rotation $R_i$ of small angle $\theta_i$ on each of the qubits at each clock cycle:
 \begin{equation}
R_i = 
\begin{pmatrix}
	    c_i & - s_i  \\
    		s_i &   c_i  \\
\end{pmatrix}
\end{equation}
with $c_i = \cos\theta_i/2$ and $s_i = \sin\theta_i/2$. The typical value of each angle is parametrized by a small angle $\eta$ to which we refer as the "precision" of the qubit: $\theta_i \sim \eta$.

We suppose that we are initially in $|\Psi_L\rangle$ and study the effect of one round of precision errors. We arrive at
$|\Psi_1\rangle$ (assuming for concision the repetition code with $N_c=3$),
\begin{eqnarray}
& &|\Psi_1\rangle = 
(\alpha c_1c_2c_3 - \beta s_1s_2s_3) |000\rangle  \\
&+& (\alpha s_1 c_2 c_3 + \beta c_1s_2s_3) |100\rangle 
+ (\alpha c_1 s_2 c_3 + \beta s_1c_2s_3) |010\rangle \nonumber\\
&+& (\alpha c_1 c_2 s_3 + \beta s_1s_2c_3) |001\rangle 
+ (\alpha c_1 s_2 s_3 - \beta s_1c_2c_3) |011\rangle \nonumber\\
&+& (\alpha s_1 c_2 s_3 - \beta c_1s_2c_3) |101\rangle 
+ (\alpha s_1 s_2 c_3 - \beta c_1c_2s_3) |110\rangle \nonumber\\
&+& (\alpha s_1 s_2 s_3 + \beta c_1c_2c_3) |111\rangle \nonumber
\end{eqnarray}
We now apply the QEC protocal and measure the stabilizers $g_1$ and $g_2$. Several results can occur. With a large probability 
$P \equiv |\alpha c_1c_2c_3 - \beta s_1s_2s_3|^2 + |\alpha s_1 s_2 s_3 + \beta c_1c_2c_3|^2 \sim 1$, we obtain no error (syndrome $+1,+1$) and the measurement projects the state directly back to the logical state. We arrive at
$|\Psi_2\rangle =\left[(\alpha c_1c_2c_3 - \beta s_1s_2s_3) |000\rangle \right.
+ \left. (\alpha s_1 s_2 s_3 + \beta c_1c_2c_3) |111\rangle \right]/\sqrt{P}$.
Since $\eta \ll 1$, we find that 
$|\Psi_2\rangle \approx  (\alpha  - \beta \eta^3/8) |000\rangle
+ (\beta  + \alpha \eta^3/8) |111\rangle$, i.e. instead of the original precision $\eta$ on the physical qubit, the logical one has a much better precision $\eta_L \sim\eta^3$. A second possible outcome of the stabilizer measurement is to observe a faulty syndrom. For instance,
with probability $P' \equiv |\alpha s_1c_2c_3 + \beta c_1s_2s_3|^2 + |\alpha c_1 s_2 s_3 - \beta s_1c_2c_3|^2 \sim \eta^2$ one gets the syndrom $(-1,+1)$ and
$|\Psi_{2'}\rangle \sim  (\alpha  + \beta \eta/2) |100\rangle
+ (\beta  + \alpha \eta/2) |011\rangle$ which after applying $X_1$ returns to the code: 
$|\Psi_{3'}\rangle \sim  (\alpha  + \beta \eta/2) |000\rangle
+ (\beta  + \alpha \eta/2) |111\rangle$. In this case, the error on the logical qubit is of the same order $\eta_L \sim\eta$ as on the original physical qubits. However, since this occurance is rare $P' \sim \eta^2$, in average the errors accumulate much more slowly. 
Generalizing to arbitrary $N_c$, we find that the logical precision after the measurement is $\eta_L \sim \eta^{N_c-2n}$ when $n$ errors are detected, which happens with probability $\eta^{2n}$. In particular, complete logical failure $\eta_L \sim 1$ happens with probability
$\eta^{N_c}$.
  
Several important conclusions can be drawn from the above analysis. 
(1) First and above, the main role of the measurement of the stabilizers is not the actual value of the syndroms  but the projection that is associated with the measurement. Even if one always find all syndroms to be $+1$ (i.e. "no error"), the role of the projection is crucial to reduce the loss of precision to $\eta_L \sim \eta^{N_c}\ll\eta$.
(2) It can be tempting to identify the parameter $p$ of the discrete error model with $p\sim \eta^2$ \cite{terhal2015,nielsen2000}. Indeed within the precision error model, the probability to get one error is  $p\sim \eta^2$ and the probability for total failure of the
logical qubit leads to the correct scaling $p_L \propto p^{N_c/2}$. However there is a crucial difference between the two error models:
when there is a number $n$ of errors close to the maximum $N_c/2$ that the code can correct, the discrete error model assumes a
perfect restoration of the logical state while in the precision error model, $\eta_L$ increases as fast as without error correction, i.e. dangerously fast. 
(3) This difference - and the fact that precision errors are always present in real life - shed some light on the notion of the "depth" of a circuit (and the associated concept of computing volume \cite{moll2017}), i.e.
how many operations one can perform on a qubit before failure. In the discrete error model, the depth is independent of the
type of calculation that one wants to perform ($\sim 1/p_L$). However, in actual devices, it will strongly depend on the effective precision needed at the end of the calculation. For instance the quantum fourrier transform algorithm \cite{shor1994} constructs an interference pattern where a very large number of unwanted states interfere destructively while the desired solution interfere constructively. Achieving such an interference requires a very high accuracy on the relative phases between the qubits.  
In evaluating a computing capability, one should therefore assess what is the needed accuracy at the end of the calculation (when the data qubits are measured) and this accuracy strongly depends on the algorithm.
(4) The community usually describes the accuracy with which a gate is performed in terms of $p$, not $\eta$. This implies for instance that
a one qubit gate performed with $99\%$ fidelity corresponds to a precision $\eta \approx 0.1$, i.e. 6 degrees.
(5) If one slightly complicates the above picture and applies the precision errors also on the ancilla qubits (i.e. in between all
gates of the circuit shown in Fig.\ref{fig:surface}d and also in between the measurements of two stabilizers), the syndrom analysis becomes more complex as one can gets wrong values of the syndroms; one must perform several measurements of the stablilizers before being able to decide on the presence of error. However, in contrast to what we wrote in an earlier version of this manuscript \cite{waintal2017}, 
the procedure still works and produces the same increase of logical precision as described above.

{\it Silent stabilizer failure error}. Hence, precision is maintained by the constant measurement of the stabilizers. Let us now examine what happens if the measurement of {\it one single } stabilizer does not occur for a few clock cycles. We further suppose that this failure is "silent", i.e. not revealed by the measurement of the corresponding ancilla. We stress that we define the stabilizer failure as the {\it absence} of measurement which is drastically different from a measurement that gives a wrong result. We will discuss below possible mechanisms that can produce such a silent stabilizer failure, but we can already note that it must happen with a finite probability $p_s$ per stabilizer since it is a local error that affects a small part of the circuit - there is always a finite probability that a small part of the circuit does not function properly. The probability to get at least one such stabilizer failure on the logical qubit is thus
$1 - (1-p_s)^{N_c-1} \sim N_c p_s$ and increases linearly with the size of the code. We argue that silent stabilizer failure errors generically leads to a failure of the logical qubit hence that that there exist a maximum value of $N_c$ above which increasing the size of the code deteriorates the logical qubit.

During the stabilizer failure, the size of the "logical" Hilbert space increases by a factor two since there is one less stabilizer 
measured. In other word, the stabilizer failure temporarily creates another "logical" qubit that we will call (temporary) silent qubit. By 
construction the silent qubit is insensitive to the measurement of the $N_c-2$ functional stabilizers. The main question to answer is
therefore will the silent qubit become entangled (hence eventually corrupt) with the logical qubit. The logical failure will happen from a combination of two mechanisms: imprecision errors as described above and another ingredient from QEC theory - fault tolerant operations -
that we now introduce. 
Indeed so far, we have shown that QEC is capable of maintaining
a certain accuracy of the logical qubit despite the presence of, say, precision errors. "Fault tolerant operations" adds the mean to manipulate the logical qubits (for instance perform a Control-Not operation between two logical qubits) within the same accuracy. By construction, these operations need to cross the large distance between the logical states. Without them, the logical qubits are essentially useless.

\begin{figure} 
\includegraphics[width=7cm]{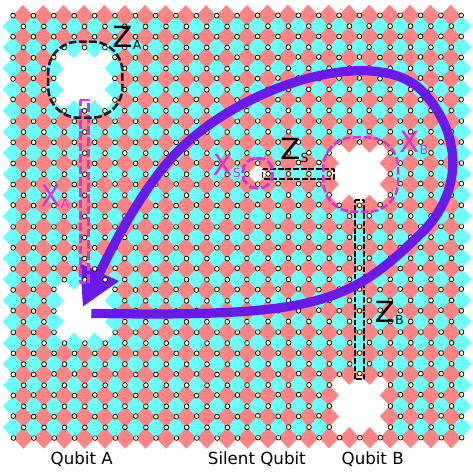} 
\vskip -0.5cm
\caption{Silent stabilizer failure. In this surface code, the logical qubits A and B are created by making two large holes (per qubit) in the code. A hole simply corresponds to not measuring the corresponding stabilizers. A Control-Not gate between A and B is performed by braiding the lower hole of A in between the two holes of B
as indicated by the blue arrow. The failure of a single stabilizer (small hole on the left of the upper hole of qubit B) creates a silent logical qubit. The dashed lines enclose the data qubits that enter in the definition of different logical operators, see text.
\label{fig:surface2} } 
\end{figure}

To be specific, we now focus on the surface code \cite{fowler2012b}. 
Fig.\ref{fig:surface2} illustrates logical qubits that are built by making pairs of holes in the surface code. 
A hole corresponds to
a bunch of stabilizers and data qubits that are (momentarily) diregarded. 
The role of $N_c$ in the repetition code is now held by the "distance of the code", i.e the number of qubits that separates the two holes
(or the number of qubits needed to make a loop around a hole, whichever of the two is the smallest).
The logical X operator of qubit A is given by $X_{\rm A} =\prod_{i\in \mathcal{A}} X_i$ where the data qubit $i$ belongs to the data set $\mathcal{A}$ inside the pink dashed line that links the two holes of qubit A, see Fig.\ref{fig:surface2}. A similar definition holds for
the $Z_A$ operator (the set loops around the upper hole of qubit A) and for qubit B (the role of X and Z is exchanged with respect to A). 
Now we suppose the silent stabilizer failure as drawn on the figure by a small hole.  This hole corresponds to an unwanted extra logical "silent" qubit. The logical Z operator $Z_{\rm S} =\prod_{i\in \mathcal{C}} Z_i$ of the silent qubit is the product of the $Z_i$ for the data set $\mathcal{C}$ of qubits enclosed in the corresponding black dashed line on Fig.\ref{fig:surface2} while its logical X operator 
$X_{\rm S} =\prod_{i\in \mathcal{C'}} X_i$ corresponds to the qubits $\mathcal{C'}$ enclosed by the pink dashed line. 
$X_S$ corresponds in fact to the missing stabilizer so that initially, the silent qubit is in a state $|+\rangle = (|0\rangle+|1\rangle)/2$. By construction, the distance (number of data qubit) of $X_S$ is $4$ while the distance of $Z_S$ is $1$ or higher ($4$ in the figure). In the case considered here, both distances are small so that precision errors will quickly build up and the silent qubit ends up in an arbitrary state on the Bloch sphere, irrespectively of the distance of the code itself. 
Logical operations provide an efficient mechanism that entangle our -now random - silent qubit with the rest of the code, hence 
jeopardizing logical qubits.
In Fig.\ref{fig:surface2}, a Control-Not gate between A and B is performed by "braiding" \cite{fowler2012b}, i.e. moving the lower hole of qubit A in a loop around the upper hole of B as indicated by the blue arrow in Fig.\ref{fig:surface2}. 
Upon braiding Qubit A with Qubit B, one also braids A with the silent qubit which creates entanglement at the logical level between S and A. This is fatal to qubit A.
We see that it is sufficient that a single  stabilizer failure occurs for the duration of one logical operation to produce an irreversible logical failure, irrespectively of $N_c$. 

{\it Examples of physical mechanisms for silent stabilizer failure.} 
Let us focus on two important physical qubits implementations, the superconducting transmon \cite{koch2007} and the semi-conducting singlet-triplet double quantum dot qubit\cite{loss1998}. In both instances, the qubits are operated using carefully shaped microwave pulses
whose frequency are resonant with a given transition. Suppose now that a classical two level system (for instance a trapped charge that can choose between two positions), sits close to an ancilla qubit. When this two level system changes its position (which may happen on any time scale from sub $\mu s$ to hours), it produces a change of the 
energy of the ancilla. As a result, the microwave pulses are no longer resonant and the pulses which are supposed to perform the Control-Not gates of Fig.\ref{fig:surface}d do, in fact, nothing to the qubit state. When the ancilla is measured, it does not perform the measurement of the stabilizer.
Note that in the language of QEC, this error (correlated failure of four Control-Not gates for several cycles) is known as a correlated error. However, it is caused by a single physical error, the "correlation" only arises from the mathematical modeling.
A second mechanism involves the so called "leakage" errors, i.e. errors that leave the code outside of its computational space.
For instance, the superconducting transmon is built out of a slightly anharmonic oscillator. When one applies a $X$ gate on a
$|0\rangle$ state, there is a small (measured in the $10^{-5}$ range \cite{chen2016}) probability that the qubit ends up not in $|1\rangle$
but in a state  $|2\rangle$ of higher energy where the qubit can stay for some clock cycles and cannot be operated. 
Singlet-triplet qubits can work in regimes where the ground state does not belong to the computational state, so that a simple relaxation can create a leakage error. 

{\it The error-by-error approach.} The strength of QEC arises from the fact that the failure of a logical qubit requires the failure of many data qubits, hence the exponential precision usually advocated. 
We have argued that this strength is also an important weakness since the data qubits are associated with as many stabilizers and the failure of a single stabilizer is sufficient to provoke the failure of the code. 
While the generic class of (silent) stabilizer failure is intrinsically fatal to the code, particular {\it instances} of this class may be faught by modifying the code \cite{aliferis2007,Silva08,Stace09,Ghosh2013,fowler2013,mehl2015,Fowler14,Hutter14} in order to monitor the corresponding error.
After such as fix, one is left with a more complex code (with more operations and/or ancillas) hence a degraded threshold. On the other hand a source of stabilizer failure has been removed hence the corresponding probability $p_s$ gets smaller, yet always finite. $p_s$ is now dominated by another error that must be identified and eventually mitigated in an error-by-error manner. 
This process is very different from simply scaling up the surface code to a larger surface. Note that we have emphasized the silent stabilizer failure error class since it is always fatal hence easier to discuss. 
A stabilizer failure that is non-silent is not necessarily fatal but might be an important hazard in practice. 
First, it effectively divides the distance of the code by 2 (in average) hence the associated loss in precision due to precision errors. 
Second, the presence of the uncontroled logical qubit puts strong restrictions on the braiding operations that can be performed, very possibly blocking the entire calculation. 
Let us end this discussion with an example of those mitigating solutions that may be attempted to detect a silent stabilizer failure. 
One could, for instance,  modify the stabilizer measurement by preparing the ancilla in a state which is not always $|0\rangle$ but alternates between $|0\rangle$ and $|1\rangle$, hence alternating syndroms $+1$,$-1$. Such a scheme might be able to detect a simple ancilla leakage, but does not work against the fluctuator discussed in the preceeding section. One could generate the alternating syndrom by adding $X$ operators on some of the data qubits during the syndrom measurement protocal (at the cost of adding extra operations hence noise in the process). Mitigating methods like that may work for detecting the stabilizer failure. If such a syndrom failure can be detected, additional steps much be taken to fix it since the code should not stay with the associated extra logical (formerly silent) qubit. These steps require either additional redondant hardware (extra ancillas) or a restructuration of the code (turning data qubits into ancilla for instance). One could imagine rarer errors (such as a direct coupling between ancillas or an indirect one mediated by a microwave pulse)
that would not be detected by the above scheme and that would require additional modifications of the code to be detected. In short, errors that affect the measurement of the stabilizers must be faught one by one in an iterative process of identification and code modification. This is in sharp contrast with the more standard type of errors that are handled by scaling.

{\it To summarize, } the global probability for a quantum calculation to fail in the plain surface code reads to leading order,
\begin{equation}
p_{\#} \sim N_{\#} N_L \left[ (p/p_{\rm th})^{\sqrt{N_c}/2} + p_s N_c \right]
\end{equation}
where $N_L$ is the number of logical qubits involved in the calculation. Following the example of
the middle size quantum computer of $N_L = 2\times 50$ logical qubits considered in \cite{reiher2017} ($N_\#> 10^{15}$, $N_c>1000$), the probability
for fatal errors must be extremely low $p_s < 10^{-20}$ for a calculation to be possible. Actual values of $p_s$ are not known, since they are currently hidden by the larger rate $p$ of correctable errors, but based on the mechanisms outline above $p_s \sim 10^{-5}$ or perhaps higher is plausible. Lowering the rate $p_s$ cannot be done by simply scaling up $N_c$. It implies a tedious error-by-error approach where very rare hidden errors must first be identified and then the QEC protocol must be complexified (at the cost of increasing $p/p_{\rm th}$) for these errors to become correctable. The literature contains many such proposals for various sort of errors \cite{aliferis2007,Silva08,Stace09,Ghosh2013,fowler2013,mehl2015,Fowler14,Hutter14}.
However, for any given architecture there will always be a mechanism that set the ultimate limit that can be achieved. This ultimate limit is the "computing power horizon" of a given technology. At a time where many different
quantum computing platforms are developped by many academic and even industrial groups, we advocate that identifying the mechanisms responsible for the (present) horizons of these technologies is a key strategic element.

\subsection*{Acknowledgment} Warm thanks to M. Sanquer who
pointed Ref.\cite{dyakonov2014} to us which triggered the present study.
The hospitality of the National University of Singapore where this work was initiated during
the French-Singapore QuESTs symposium is also acknowledged. We thank A. Fowler, T. Meunier, JM Gerard, N. Mahrour, M. Houzet, 
P. Bertet, N. Roch and M. Mirrahimi for useful discussions or correspondance.

\bibliographystyle{apsrev}
\bibliography{qu_err_corr.bib}

\end{document}